\documentclass[journal]{IEEEtran}
\usepackage[pdftex]{graphicx}
\usepackage{cite}
\usepackage{amsmath,amssymb,amsfonts}
\usepackage{algorithmic}
\usepackage{textcomp}
\usepackage{color,soul}
\usepackage{fancyhdr}
\renewcommand{\thispagestyle}[1]{} 

\begin{document}

\renewcommand{\thispagestyle}[1]{} 

\pagestyle{fancy}

\title{Cryogenic Subthreshold Swing Saturation in FD-SOI MOSFETs described with Band Broadening}

\author{H.~Bohuslavskyi, A.~G.~M.~Jansen, S.~Barraud, V.~Barral, M.~Cass\'e, L.~Le~Guevel,  X.~Jehl, L.~Hutin, B.~Bertrand, G.~Billiot, G.~Pillonnet, F.~Arnaud, P.~Galy, S.~De~Franceschi, M.~Vinet, and M.~Sanquer 
\thanks{This work was supported by the EU Research and Innovation program Horizon 2020  under grant agreement No 688539 MOSQUITO.}
\thanks{H.~Bohuslavskyi, S.~Barraud, V.~Barral, M. ~Cass\'e, L.~Le~Guevel, L.~Hutin, B.~Bertrand, G.~Billiot, G.~Pillonnet, and M.~Vinet are with CEA, LETI, Minatec Campus, F-38054 Grenoble, France.} 
\thanks{A.~G.~M.~Jansen, X.~Jehl, S.~De~Franceschi, and M.~Sanquer are with Univ. Grenoble Alpes, CEA, INAC-PHELIQS, F-38054 Grenoble, France.} 
\thanks{F.~Arnaud and P.~Galy are with STMicroelectronics, 850 rue J. Monnet, 38920 Crolles, France.} 
\thanks{Corresponding authors: marc.sanquer@cea.fr and louis.jansen@cea.fr}}

\maketitle

\begin{abstract}
\boldmath
In the standard MOSFET description of the drain current $I_{D}$ as a function of applied gate voltage $V_{GS}$, the subthreshold swing $SS(T)\equiv dV_{GS}/d\log I_{D}$ has a fundamental lower limit as a function of temperature $T$ given by $SS(T)  = \ln10~k_BT/e$. 
However, recent low-temperature studies of different advanced CMOS technologies have reported $SS$(4~K or lower) values that are at least an order of magnitude larger. 
Here, we present and analyze the saturation of $SS(T)$ in 28~nm fully-depleted silicon-on-insulator (FD-SOI) devices for both n- and p-type MOSFETs of different gate oxide thicknesses and gate lengths down to 4~K. 
Until now, the increase of interface-trap density close to the band edge as temperature  decreases has been put forward to understand the saturation.
Here, an original explanation of the phenomenon is presented by considering a disorder-induced tail in the density of states at the conduction 
(valence) band edge for the calculation of the MOS channel transport by applying Fermi-Dirac statistics.
This results in a subthreshold
$I_{D}\sim e^{eV_{GS}/k_BT_0}$ for $T_0=35$~K with saturation value $SS(T<T_0) =  \ln 10~k_BT_0/e$.
The proposed model adequately describes the experimental data of $SS(T)$ from 300 down to 4~K using $k_BT_0 \simeq 3$~meV for the width of the exponential tail and can also accurately describe $SS(I_{D})$ within the whole subthreshold region.
Our analysis allows a direct determination of the technology-dependent band-tail extension forming a crucial element in future compact modeling and design of cryogenic circuits. 
\unboldmath
\end{abstract}


\section{Introduction}
The development of electronic circuits at cryogenic temperatures (4~K or even lower) has great importance for a large spectrum of applications such as high-performance classical computing, cryogenic sensors and detectors, space electronics, low power neuromorphic circuits, and quantum computing \cite{Gutierrez2001,Reilly2015,Homulle2018,Schneider2018}.
The nowadays progress in the realization of quantum bit (qubit) systems at low temperatures has proven the necessity of having nearby cryogenic electronics to enable fast and efficiently-controlled manipulation and read-out of a large number of qubits \cite{Reilly2015,Patra2018,Vandersijpen2017}. 
In this respect, the recent demonstrations of silicon qubits  could be combined with state-of-the-art CMOS electronics \cite{Muhonen2014,Veldhorst2015,Mauraud2016}. 

The 28nm fully-depleted silicon-on-insulator (FD-SOI) MOSFETs with undoped channel have numerous advantages for low-temperature applications compared to Si bulk transistors, such as the reduced impact of dopant freeze-out, reduced variability, tuning of threshold voltage thanks to back-biasing, higher mobility, and quasi-ideal electrostatic control \cite{Jacquet2014,Doris2016}.

The present study focuses on the exponential gate-voltage ($V_{GS}$) dependence of the subthreshold drain current $I_{D}$ of a MOSFET  as a function of the temperature $T$ captured by the subthreshold swing $SS(T) \equiv dV_{GS}/d\log I_{D} = m \ln 10~ k_BT/e$ \cite{Sze-book} 
($SS(300~K)=60$~mV/dec for $m=1$), 
where $k_B$ is the Boltzmann constant and $e$ the absolute elementary charge.
The factor $m=(C_{ox}+C_{it})/C_{ox} \geq 1$ takes into account the capacitance of the interface traps $C_{it}$ with respect to the geometric gate capacitance $C_{ox}$ 
(in our case of FD-SOI, the depletion capacitance can be neglected). 
The very small value of $SS$ in the (sub)K-range results in an ideal $I_{D}(V_{GS})$ switch leading to a high on/off current ratio and low power dissipation in the stand-by regime.  

Cryogenic investigations of $SS(T)$ in advanced CMOS devices reveal at least one order of magnitude larger values at 4 K compared with the expected value $SS(4K)=0.79$~mV/dec for $m=1$. For bulk MOSFETs, typical values for SS(4~K) range from 30 to 10~mV/dec \cite{Honamura1986,Gutierrez2001,Incandela2017,Homulle2018}. 
Even though an effective operation of different SOI technologies at cryogenic temperature has been demonstrated, the reported SS values turned out to be as high as 7~mV/dec 
at 4K~\cite{Roche2012,Shin2014,Bohuslavskyi2017,Beckers2018a},
even at sub-1K temperatures \cite{Galy2018}.

Common explanations correlate the $SS(T)$ dependence in bulk Si transistors to an important increase of the density of interface traps $D_{it}$ close to the band edges~\cite{Hafez1990}.
Similarly, for planar SOI devices, the increase of  $D_{it}$ has been demonstrated to extend more than 100~meV inside the band gap using the spectroscopic charge-pumping  technique~\cite{Casse2010}. However, especially at the lowest temperatures, analyzing the saturation with a strongly temperature dependent increase of $m$ via $C_{it}=e^2D_{it}$ leads to an estimate of unrealistic $D_{it}$ values that are even larger than the silicon density of states of free carriers~\cite{Galy2018}. 
More recently, a constant contribution to $SS(T)$ has been derived at 4~K by modeling the thermal occupation of the interface-trap distribution~
\cite{Beckers2018a,Beckers2018b,Beckers2018c,Beckers2018d}
which needs temperature-specific modeling for an application to all temperatures. 

In order to explain the cryogenic saturation of $SS$, we propose a new approach introducing a disorder-induced exponential tail in the density of states (DOS) for the calculation of the subthreshold charge-carrier transport. 
The model is validated on the experimental data of long- and short-channel MOSFETs with different oxide thicknesses in terms of both $SS(T)$ and $SS(I_{D})$ dependences from $300$~K down to $4.3$~K.

\section{Experiment}

Thin (GO1) and thick (GO2) gate oxide low-threshold-voltage (LVT) FD-SOI transistors were fabricated with a 
gate-first high-$\kappa$ metal gate by STMicroelectronics
on 300~mm (100) SOI wafers with a buried oxide thickness of 25nm \cite{Planes2012,Jacquet2014}. 
The equivalent oxide thickness (EOT) is 1.55~nm for GO1 and 3.7~nm for GO2.  
The 7~nm-thick channel is undoped. 
 
Both n-type and p-type  long-channel GO2 (gate length $L=0.15, 2$~$\mu$m and width $W=2$~$\mu$m) and n-type short-channel GO1
($L=28, 34$~nm and $W=80, 210$~nm) transistors were investigated.
P-type short-channel devices couldn't be analyzed for the $SS(T)$-dependence because of oscillatory variations in $SS(I_{D})$ below threshold at cryogenic temperatures. This results from the enhanced boron diffusion from the source/drain regions affecting subthreshold current at low $V_{DS}$ 
(see~\cite{Wacquez2010,Bohuslavskyi2017}).

The transistors cleaved from a wafer were mounted to the sample holder of a cryogenic probe station equipped with 4 adjustable contact needles connected to Source/Measurement Units. 
Then, they were cooled down under continuous He flow with temperature regulation between 4 and 300~K. 
The data acquisition was done using a parameter analyzer (HP 4155A).

Fig.~1a shows $I_{D}(V_{GS})$ at temperatures between 300 and 4.3~K for the n-type long- and short-channel devices
at $|V_{DS}| = 50$~mV and back-gate voltage $V_{BACK} = 0$~V. 
Both cases reveal a classical, oscillation-free $I_{D}(V_{GS})$ down to the lowest temperatures.
The saturation of $SS(T)$ below about 40~K can be clearly seen from the $SS(I_{D})$ data as illustrated in Fig.~1b. 
The same trend holds for p-type long-channel and n-type short-channel devices.

In Fig.~1c, $SS(T)$ from 300~K down to roughly 40 K follows the expected dependence $m_{1,2}\ln10~k_BT/e$ with $m_1=1.14$ for the long devices and $m_2=1.23$ for the short device. 
The slightly higher $SS$ (described by $m_1 = 1.14$)
for the long devices is explained by the presence of interface traps ($C_{it}$), 
and the larger $m_2 = 1.23$ follows from additional electrostatic short-channel effects~\cite{Lundstrom2018}. 
$SS(4.3~{\rm K})$ saturates at 7.3~mV/dec for n-type long-channel, 7.4~mV/dec for p-type long-channel, and 7.7~mV/dec for n-type short-channel.

\begin{figure}[t!]
\centering
\includegraphics[width=1\linewidth]{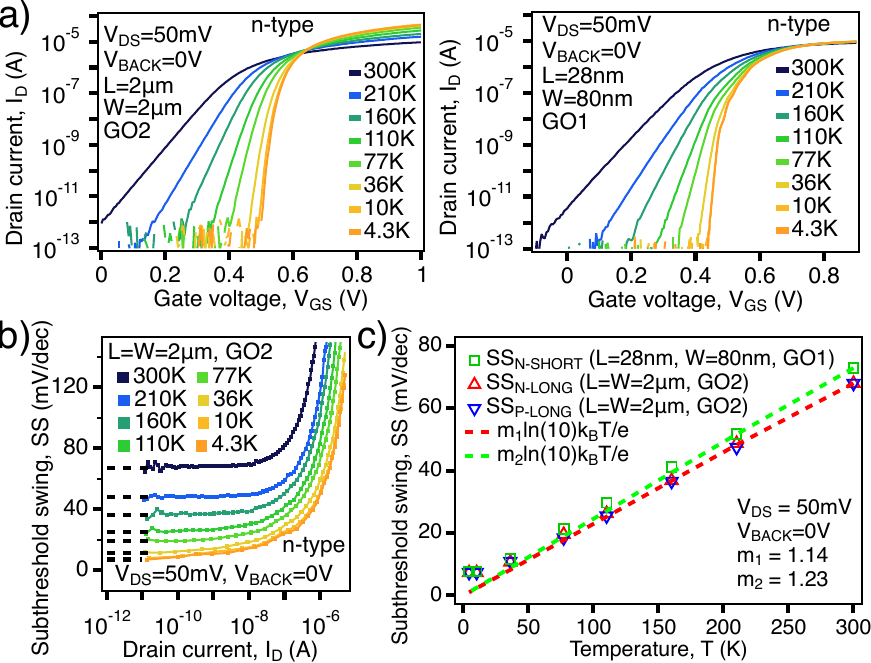}   
\caption{
a) $I_{D}(V_{GS})$ characteristics  at different temperatures recorded at $V_{DS}=50$~mV. Left and right panel show the case of n-type long- and short-channel MOSFETs. 
b) $SS(I_{D})$ for the n-type long-channel. 
c) $SS(T)$ obtained by extrapolating $SS$ at $I_{D} = 10^{-11}$~A for n- and  p-type long-channel ($SS_{N,P-LONG}$) and n-type short-channel ($SS_{N-SHORT}$) transistors. Dashed lines stand for the expected linear temperature dependence $m_{1,2} \ln 10~k_BT/e$ with $m_1 = 1.14$ and $m_2 = 1.23$.
}
\end{figure}

\section{Model description and discussion}

The diffusive subthreshold transport is proportional to the density $n$ of the mobile charge carriers in the channel assuming a constant diffusion constant (mobility)~\cite{Sze-book}.
Using the Fermi-Dirac statistics for the occupation of electron states~\cite{Lundstrom2018,Ma2015}, 
here for the n-type case,  
$n$ can be expressed  as a function of the semiconductor potential $\Psi_s$ via
\begin{equation}
n(\Psi_s) = \int_{-\infty}^{\infty} f(E)N^{2D}_c(E) dE ,
\end{equation} 
with the Fermi function $f(E)=1/({\rm e}^{(E - E_F)/k_BT}+1)$ and a step function for the two-dimensional DOS $N^{2D}_c(E)$ from zero to $N^{2D}_c = g_v m^* /\pi \hbar^2$ at the band edge $E_c =E_c^0 -e\Psi_s$. 
For the flat band condition with $\Psi_s=0$, the Fermi energy $E_F$ is taken at mid-gap with $E_c^0 = 0.55$~eV 
(considering a temperature-independent energy gap 1.1~eV for Si).  
Other parameters are the valley degeneracy $g_v=2$ and the effective mass $m^* = 0.19~m_0$ (free-electron mass $m_0$). 
Finally, the equilibrium electron density $n(\Psi_s)$ can be 
transposed to $n(V_{GS})$ using 
$V_{GS} = \Psi_s + n(\Psi_s)e/C_{ox}$ \cite{Lundstrom2018}, for the sum of the semiconductor potential $\Psi_s$ and the voltage  drop $ne/C_{ox}$ over the geometric gate capacitance $C_{ox} = k\epsilon_0/t_{EOT}$~[F/m$^2$] supposing $m=1$. $\epsilon_0$ is the free-space permittivity, $k=3.9$ the relative dielectric constant of SiO$_2$, and $t_{EOT}=3.7$~nm the equivalent oxide thickness in case of GO2.   
 
The calculated $n(V_{GS})$ data for a sharp band edge reveal the standard exponential dependence 
$I_{D} \sim e^{eV_{GS}/k_BT}$, confirming the linear temperature dependence $SS(T) = \ln 10~k_BT/e$. 
However, only using the Fermi-Dirac statistics is not enough to explain
the experimentally observed saturation at low temperatures as shown in Fig.~1c.

To describe the saturation of $SS$ at low temperatures, a broadened band edge \cite{Lifshitz1964,Gold1988} was added to the DOS in the form 
$N^{2D}_{c} {\rm e}^{(E - E_c)/k_BT_0}$ for $E<E_c$  (inset in Fig.~2c). 
The parameter $k_BT_0$ quantifies the extent of the exponential tail resulting from, e.g., crystalline disorder, residual impurities, and strain, surface roughness, etc.
Assuming a proportionality between $I_{D}(V_{GS})$ and $n(V_{GS})$, the calculated $I_{D}(V_{GS})$ is shown in Fig.~2a for different temperatures with $T_0=35$~K (resulting in $k_BT_0=3$~meV). 
A saturation value $SS(T \leq T_0) = \ln 10~k_B T_0/e = 6.9$~mV/dec is obtained for $m = 1$ (see Fig.~2b).
The $3$~meV tail was determined empirically to describe the experimental $SS$(4.3~K) of 7-8~mV/dec.

\begin{figure}[t!]
\centering
\includegraphics[width=1\linewidth]{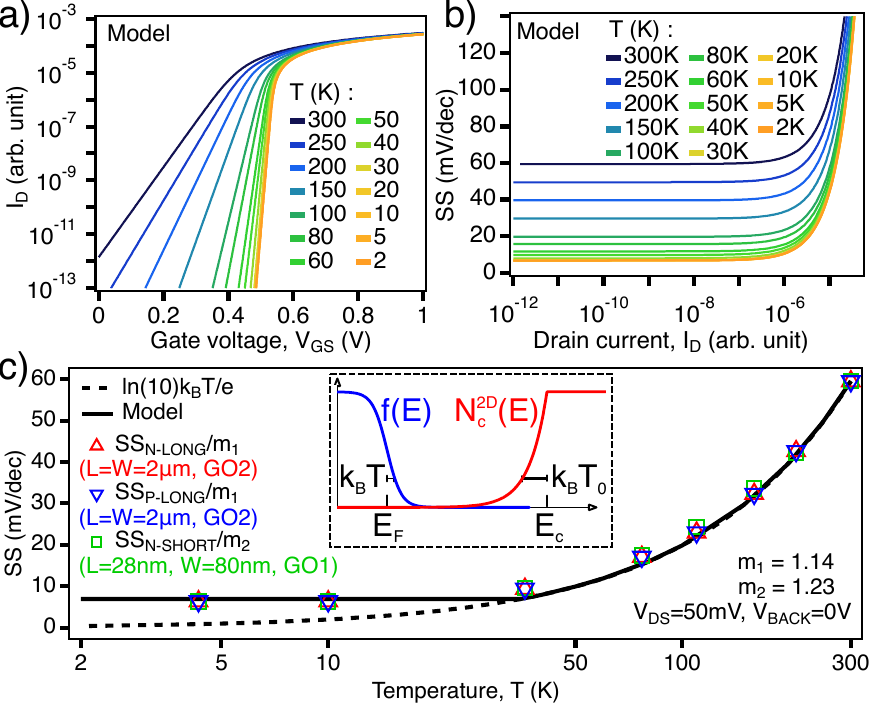}
\caption{
a) Calculated $I_{D}(V_{GS})$ for the model parameters given in the main text, assuming an exponential tail below the conduction band edge with $k_BT_0=3$~meV. 
b) Calculated $SS(I_{D})$. 
c) Calculated $SS(T)$ compared to experimental data normalized by $m_{1,2}$ (symbols) and to $\ln 10~k_B T/e$. The inset shows a schematic representation of the  exponential tails of the Fermi distribution function $f(E)$ for $E>E_F$ and of the DOS $N_c^{2D}(E)$ for $E<E_c$. The product of $f(E)$ and $N_c^{2D}(E)$ for $T < T_0$ gives the distribution of occupied states with a maximum around $E_F$. 
}
\end{figure}

To compare with the model, 
the saturation values $SS(T)$ in the weak inversion measured at $I_{D}=10$~pA are plotted in Fig.~2c after normalization with the corresponding $m_{1,2}$ (for the values, see Fig.~1c). 
It should be noted that the chosen $k_BT_0$ gives a good description of the experimental data for all studied MOSFETs.
The obtained exponential extent for the band tail is comparable to that of 2-10~meV probed with Electron Spin Resonance on Si MOSFETs in
\cite{Jock2012}. 

 FD-SOI cryogenic back-biasing was already demonstrated to be efficient down to 4~K \cite{Bohuslavskyi2018}.  
By using forward back-biasing (FBB), the conductive channel can be displaced towards the Si-BOX interface. Therefore, if the increase of $D_{it}$ was responsible for $SS(T)$ saturation, one would expect a significant difference in $SS(I_{D})$ profiles. 
However, the experimental data in Fig. 3a and 3b for an n-type device at 4.3~K reveal that $SS(I_{D})$ curves hardly change for $V_{BACK}$ up to 3~V, implying that the edge-broadened DOS cannot be explained with just $D_{it}$ at the Si-SiO$_2$ interface. 

 At the lowest temperatures, the measured $SS(I_{D})$ characteristics reveal an increased gate-voltage dependence (Fig.~1b) as compared to the constant $SS(I_{D})$ from our calculations (Fig.~2b). This variation of $SS(I_{D})$ can also be modeled by including an energy  dependence $m(E)$ in the relation $V_{GS}(\Psi_s) = m(\Psi_s)\Psi_s + n(\Psi_s)e/C_{ox}$, similarly to the description of the interface traps below the band edge with the additional capacitance $C_{it}$ in the introduction. 
In Fig.~3c the calculated $SS(I_{D})$ is shown for an exponential dependence $m(E) \sim e^{(E-E_c)/E_m}$.  A good agreement with the experimental data is found for a variation of $m(E)$ from 1.14 to 1.34 with an empirically-determined energy range $E_m = 10$~meV below $E_c$.
Regarding the physical reasoning behind the improved model which includes $m(E)$, we note that not all the states in the band tail contribute to the  transport~\cite{MottDavis1979}. 
Therefore, the $V_{GS}$-induced occupation of states in the band tail influences both the subthreshold current due to mobile states and the efficiency of gate control via $m(\Psi)$ because of trapped states.
In our modeling, each of these contributions has its characteristic extent in energy below $E_c$. 

\begin{figure}[t!]
\centering
\includegraphics[width=1\linewidth]{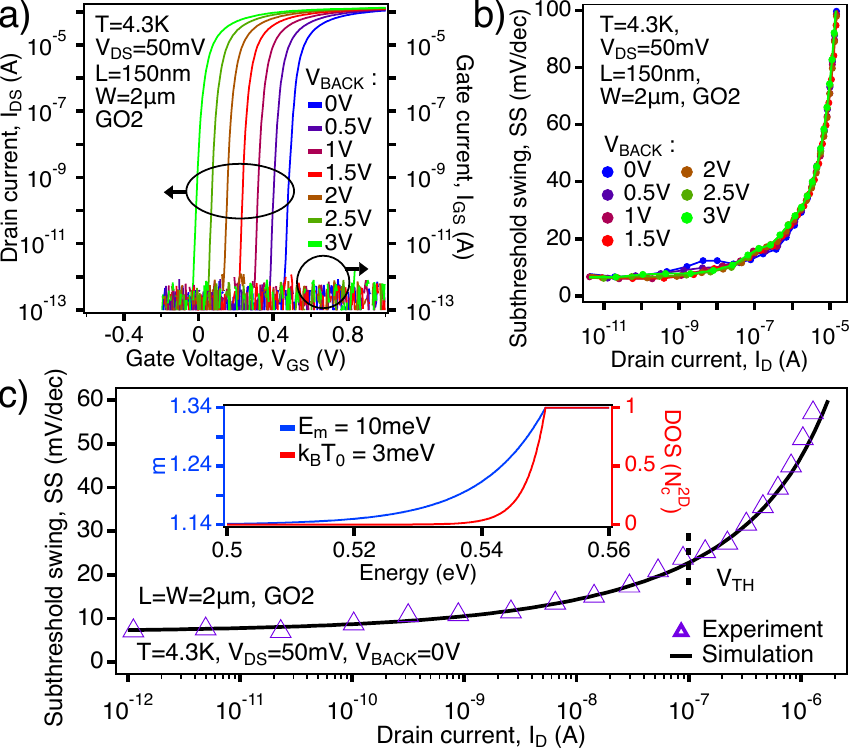}  
\caption{
a) $I_{D}(V_{GS})$ and $I_{G}(V_{GS})$ for different $V_{BACK}$ and b) extracted $SS(I_{D})$ for the same device at $T = 4.3$~K. 
Almost the same $SS(I_{D})$ profile is observed when the conductive channel is pulled towards the Si-BOX interface under FBB. 
c) $SS(I_{D})$ profile with $SS$ ranging from 7.4~mV/dec ($I_{D}=1$~pA) to 23~mV/dec at $V_{TH}$ ($I_{D}=0.1~\mu$A) for a long-channel n-type device illustrating the increased $SS(V_{GS})$ dependence at $T = 4.3$~K.
Accounting for the exponential energy-dependent $m(E)$ and $k_{B}T_0 = 3$~meV (as shown in the inset), the refined model accurately reproduces the experimental $SS(I_{D})$ within the whole subthreshold region in addition to the $SS(T)$ dependence from $300$ down to $4.3$~K (see Fig.~2c).
}
\end{figure}

\section{Conclusion}
To explain the generally observed saturation of $SS$ at low temperature in FD-SOI MOSFETs, an exponential tail at the band edge is introduced yielding  $I_{D}(V_{GS})$ of the form $e^{eV_{GS}/k_BT_0}$ that replaces the usual temperature dependence $e^{eV_{GS}/k_BT}$ for $T < T_0$.
The determined  $T_0=35$~K holds for all measured FD-SOI devices with long- and short-channel lengths for different oxide thickness and accurately describe the $SS(T)$ from 300~K down to 4.3~K. 
In addition, we address the problem of the increased cryogenic $SS(V_{GS})$ dependence at low temperatures and successfully model a non-constant $SS(I_{D})$ profile below $V_{TH}$ by introducing an energy-dependent $m(E)$ in the gate-control efficiency.
Finally, our results indicate that the implementation of band-tail broadening could form an important technological parameter for the correct modeling of MOSFETs at low temperatures.


\begin{thebibliography}{11}
\bibitem{Reilly2015}
D.~J.~Reilly, 
"Engineering the quantum-classical interface of solid-state qubits",
\emph{Npj Quantum Information}, 
vol. 1, no. 15011,
Oct. 2015.
doi: 10.1038/npjqi.2015.11

\bibitem{Schneider2018}
M.~L.~Schneider, C.~A.~Donnelly, S.~E.~Russek, B.~Baek, M.~R.~Pufall, P.~F.~Hopkins, P.~D.~Dresselhaus, S.~P.~Benz, and W.~H.~Rippard,
"Ultralow power artificial synapses using nanotextured magnetic Josephson junctions",
\emph{Science Advances},
vol. 4, no. 1, 
Jan. 2018.
doi: 10.1126/sciadv.1701329

\bibitem{Gutierrez2001}
E.~Gutierrez-D, J.~Deen, and C.~Claeys,
\emph{Low Temperature Electronics: Physics, Devices, Circuits, and Applications},
San Diego, Academic Press, 
Oct. 2001,
pp. 135 - 240.

\bibitem{Homulle2018}
H.~Homulle, L.~Song, E.~Charbon, and F.~Sebastiano,
"The Cryogenic Temperature Behavior of Bipolar, MOS, and DTMOS Transistors in Standard CMOS,
\emph{IEEE Journal of the Electron Devices Society},
vol. 6, pp. 263-270,
2018.
doi: 10.1109/JEDS.2018.2798281

\bibitem{Patra2018}
B.~Patra, R.M.~Incandela, J.~P.~G.~van Dijk, H.~A.~R.~Homulle,
L.~Song, M.~Shahmohammadi, R.~B.~Staszewski, A.~Vladimirescu, 
M.~Babaie, F.~Sebastiano, and E.~Charbon,
"Cryo-CMOS Circuits and Systems for Quantum
Computing Applications",
\emph{IEEE Journal of Solid-State Circuits},
vol. 53, pp. 309-321,
2018. 
doi: 10.1109/JSSC.2017.2737549

\bibitem{Vandersijpen2017}
L.~M.~K.~Vandersypen, H.~Bluhm, J.~S.~Clarke, A.~S.~Dzurak, R.~Ishihara, A.~Morello, D.~J.~Reilly, L.~R.~Schreiber, and M.~Veldhorst,
"Interfacing spin qubits in quantum dots and donors - hot, dense, and coherent",
\emph{Npj Quantum Information},
vol. 3, no. 34,
Sept. 2017.
doi: 10.1038/s41534-017-0038-y15011

\bibitem{Muhonen2014}
 J.~T.~Muhonen, J.~P.~Dehollain, A.~Laucht, F.~E.~Hudson, R.~Kalra, T.~Sekiguchi, K.~M.~Itoh, D.~N.~Jamieson, J.~C.~McCallum, A.~S.~Dzurak, and A.~Morello,
 "Storing quantum information for 30 seconds in a nanoelectronic device"
\emph{Nature Nanotechnology},
vol. 9, pp. 986 - 981,
Sept. 2014.
doi: 10.1038/NNANO.2014.211

\bibitem{Veldhorst2015}
M.~Veldhorst, C.~H.~Yang, J.~C.~C.~Hwang, W.~Huang, J.~P.~Dehollain, J.~T.~Muhonen, S.~Simmons, A.~Laucht, F.~E.~Hudson, K.~M.~Itoh, A.~Morello, and A.~S.~Dzurak,
"A two-qubit logic gate in silicon",
\emph{Nature},
vol. 526, pp. 410 - 414, 
Oct. 2015.
doi: 10.1038/nature15263

\bibitem{Mauraud2016}
R.~Maurand, X.~Jehl, D.~Kotekar-Patil, A.~Corna, H.~Bohuslavskyi, R.~Lavieville, L.~Hutin, S.~Barraud, M.~Vinet, M.~Sanquer, and S.~De Franceschi, 
"A CMOS silicon spin qubit", 
\emph{Nature Communications},
vol. 7, no. 13575, 
Oct. 2016. 
doi: 10.1038/ncomms13575 

\bibitem{Doris2016}
B.~Doris, B.~De Salvo, K.~Cheng, P.~Morin, and M.~Vinet,
"Planar Fully-Depleted-Silicon-On-Insulator technologies: Toward the 28nm node and beyond",
\emph{Solid-State Electronics},
vol. 117, pp. 37-59, 
Mar. 2016. 
doi: 10.1016/j.sse.2015.11.006

\bibitem{Jacquet2014}
D.~Jacquet, F.~Hasbani, P.~Flatresse, R.~Wilson, F.~Arnaud, G.~Cesana, T.~Di Gilio, C.~Lecocq, T.~Roy, A.~Chhabra, C.~Grover, O.~Minez, J.~Uginet,
G.~Durieu, C.~Adobati, D.~Casalotto, F.~Nyer, P.~Menut, A.~Cathelin, I.~Vongsavady, and P.~Magarshack, 
"A 3 GHz dual core processor ARM Cortex TM -A9 in 28 nm UTBB FD-SOI CMOS with ultra-wide voltage range and energy efficiency optimization",
\emph{IEEE Journal of Solid-State Circuits},
vol. 49, pp. 812 - 826, 
Jan. 2014. 
doi: 10.1109/JSSC.2013.2295977

\bibitem{Sze-book}
S.~M.~Sze,
\emph{Physics of semiconductor devices (2nd Edition)}, 
New York, John Wiley \& Sons, 1981, pp. 431 - 511.

\bibitem{Honamura1986}
H.~Hanamura, M.~Aoki, T.~Masuhara, O.~Minato, Y.~Sakai, and T.~Hayashida, 
"Operation of bulk CMOS devices at very low temperatures",
\emph{IEEE Journal of Solid-State Circuits}
vol. 21, no. 3, pp. 484 - 490, 
Jun. 1986.
doi: 10.1109/JSSC.1986.1052555

\bibitem{Incandela2017}
R.~M.~Incandela, L.~Song, H.~A.~R.~Homulle, F.~Sebastiano, E.~Charbon, and A.~Vladimirescu,
"Nanometer CMOS characterization and compact modelling at deep-cryogenic temperatures",
\emph{2017 47th European Solid-State Device Research Conference (ESSDERC)},
Leuven, 2017, pp. 58 - 61.
doi: 10.1109/ESSDERC.2017.8066591

\bibitem{Roche2012}
B.~Roche, B.~Voisin, X.~Jehl, R.~Wacquez, M.~Sanquer, M.~Vinet, V.~Deshpande, and B.~Previtali,
"A tunable, dual mode field-effect or single electron transistor",
\emph{Applied Physics Letters},
vol. 100, no. 032107, 
Aug. 2012. 
doi: 10.1063/1.3678042

\bibitem{Shin2014}
M.~Shin, M.~Shi, M.~Mouis, A.~Cros, E.~Josse, G.~T.~Kim, and G.~Ghibaudo,
"Low temperature characterization of14nm FDSOI CMOS devices",
\emph{2014 11th International Workshop on Low Temperature Electronics (WOLTE)},
Grenoble, 2014, pp. 29 - 32. 
doi: 10.1109/WOLTE.2014.6881018

\bibitem{Beckers2018a}
A~Beckers, F.~Jazaeri, H.~Bohuslavskyi, L.~Hutin, S.~De Franceschi, and C.~Enz,
"Design-oriented modeling of 28 nm FDSOI CMOS technology down to 4.2K for quantum computing"
\emph{2018 Joint International EUROSOI Workshop and International Conference on Ultimate Integration on Silicon (EUROSOI-ULIS)},
Granada, 2018, pp. 1 - 4.
doi: 10.1109/ULIS.2018.8354742

\bibitem{Bohuslavskyi2017}
H.~Bohuslavskyi, S.~Barraud, M.~Cass\'{e}, V.~Barra, B.~Betrand, L.~Hutin,
F.~Arnaud, P.~Galy, M.~Sanquer, S.~De Franceschi, and M.~Vinet,
28nm Fully-Depleted SOI Technology:
"Cryogenic Control Electronics for Quantum Computing",
\emph{2017 Silicon Nanoelectronics Workshop}, 
Kyoto, 2017, pp. 143-144. 
doi: 10.23919/SNW.2017.8242338

\bibitem{Galy2018}
P.~Galy, J.~Camirand Lemyre, P.~Lemieux, F.~Arnaud, D.~Drouin, and M.~Pioro-Ladri\`{e}re,
"Cryogenic temperature characterization of a 28-nm FD-SOI dedicated structure for advanced CMOS and quantum technologies co-integration"
\emph{IEEE Journal of the Electron Devices Society},
vol. 6, pp. 594 - 600,
May 2018. 
doi: 10.1109/JEDS.2018.2828465

\bibitem{Hafez1990}
I.~M.~Hafez, G.~Ghibaudo, and F.~Balestra, 
"Assessment of interface state density in silicon MOS transistors at room, liquid nitrogen, and liquid helium temperatures"
\emph{Journal of Applied Physics}, 
vol. 4, no. 4, pp. 1950 - 1952,
Oct. 1990.
doi: 10.1063/1.345572

\bibitem{Casse2010}
M.~Cass\'{e}, K.~Tachi, S.~Thiele, and T.~Ernst,
"Spectroscopic charge pumping in Si nanowire transistors with a high-$\kappa$/metal gate",
\emph{Applied Physics Letters},
vol. 96, no. 123506, 
Mar. 2010.
doi: 10.1063/1.3368122

\bibitem{Beckers2018b}
A.~Beckers, F.~Jazaeri, and C.~Enz,
"Cryogenic MOS transistor model",
\emph{IEEE Transactions Electron Devices},
vol. 65, pp. 3617 - 3625, 
Aug. 2018.
doi: 10.1109/TED.2018.2854701

\bibitem{Beckers2018c}
A. Beckers, F. Jazaeri and C. Enz, 
"Characterization and modeling of 28-nm Bulk CMOS Technology Down to 4.2 K"
\emph{IEEE Journal of the Electron Devices Society}, 
vol. 6, pp. 1007-1018, Mar. 2018.
doi: 10.1109/JEDS.2018.2817458

\bibitem{Beckers2018d}
A. Beckers, F. Jazaeri and C. Enz, 
"Revised theoretical limit of the subthreshold swing in field-effect transistors",
\emph{arXiv:1811.09146v1 [cond-mat.mes-hall]}, 
Nov. 2018.

\bibitem{Planes2012}
N.~Planes, O.~Weber, V.~Barral, S.~Haendler, D.~Noblet, D.~Croain, M.~Bocat, P.-O.~Sassoulas, X.~Federspiel, A.~Cros, A.~Bajolet, E.~Richard, B.~Dumont, P.~Perreau, D.~Petit, D.~Golanski, C.~Fenouillet-Béranger, N.~Guillot, M.~Rafik, V.~Huard, S.~Puget, X.~Montagner, M.-A.~Jaud, O.~Rozeau, O.~Saxod, F.~Wacquant, F.~Monsieur, D.~Barge, L.~Pinzelli, M.~Mellier, F.~Boeuf, F.~Arnaud, and M.~Haond, 
"28nm FDSOI technology platform for high-speed low-voltage digital applications",
\emph{2012 Symposium on VLSI Technology (VLSIT)},
Honolulu, HI, 2012, pp. 133 - 134. 
doi: 10.1109/VLSIT.2012.6242497

\bibitem{Wacquez2010}
R.~Wacquez, M.~Vinet, M.~Pierre, B.~Roche, X.~Jehl, O.~Cueto, J.~Verduijn, G.~C.~Tettamanzi, S.~Rogge, V.~Deshpande, B.~Previtali, C.~Vizioz, S.~Pauliac-Vaujour, C.~Comboroure, N.~Bove, O.~Faynot, and M.~Sanquer,
"Single dopant impact on electrical characteristics of NMOSFETs with effective length down to 10nm",
\emph{2010 Symposium on VLSI Technology},
Honolulu, 2010, pp. 193 - 194.
doi: 10.1109/VLSIT.2010.5556224

\bibitem{Lundstrom2018}
M.~Lundstrom, 
\emph{Fundamentals of nanotransistors (Lessons from Nanoscience: A Lecture Notes Series, Vol. 6)}, 
Singapore, World Scientific Co. Pte. Ltd., 2018, pp. 143 - 181.

\bibitem{Ma2015}
N.~Ma and D.~Jena, 
"Carrier statistics and quantum capacitance effects on mobility extraction in two-dimensional crystal semiconductor field-effect transistors",
\emph{2D Materials},
vol. 2, no. 015003,
Jan. 2015. 
doi: 10.1088/2053-1583/2/1/015003

\bibitem{Lifshitz1964}
I.~M.~Lifshitz,
"The energy spectrum of disordered systems", 
\emph{Advances in Physics},
vol. 13, no. 483, pp. 483 - 536. 
Oct. 1964.
doi: 10.1080/00018736400101061

\bibitem{Gold1988}
A.~Gold, J.~Serre, and A.~Ghazali,
"Density of states in a two-dimensional electron gas: Impurity bands and band tails",
\emph{Phys. Rev. B},
vol. 37, no. 9, pp. 4589 - 4603,
Mar. 1988.
doi: 10.1103/PhysRevB.37.4589

\bibitem{Jock2012}
R.~M.~Jock, S.~Shankar, A.~M.~Tyryshkin, J.~He, K.~Eng, K.~D.~Childs, L.~A.~Tracy, M.~P.~Lilly, M.~S.~Carroll,
and S.~A.~Lyon,
"Probing band-tail states in silicon metal-oxide-semiconductor heterostructures with
electron spin resonance",
\emph{Appl. Phys. Lett.},
vol. 100, 023503, 
Jan. 2012.
doi: 10.1063/1.3675862 

\bibitem{Bohuslavskyi2018}
H.~Bohuslavskyi, S.~Barraud, V.~Barral, M.~Casse, L.~Le~Guevel, L.~Hutin, B.~Bertrand, A.~Crippa, X.~Jehl, G.~Pillonnet, A.~G.~M.~Jansen, F.~Arnaud, P.~Galy, R.~Maurand, S.~De~Franceschi, and M.~Vinet,
"Cryogenic Characterization of 28-nm FD-SOI Ring Oscillators With Energy Efficiency Optimization",
\emph{IEEE Transactions on Electron Devices},
vol. 65, no. 9, pp. 3682-3688, 
Sept. 2018.
doi: 10.1109/TED.2018.2859636

\bibitem{MottDavis1979}
N.~F.~Mott and E.~A.~Davis, 
\emph{Electronic Processes in Non-crystalline Materials (2nd Edition)},
Oxford, Clarendon Press, 1979, pp. 7 - 62. 

\end{thebibliography}
\end{document}